\documentclass[a4paper,12pt]{article}
\usepackage[reqno,fleqn]{amsmath}
\usepackage{color,graphicx}
\usepackage[T1]{fontenc}
\author{Elnaz Amirkhanlou\footnote{eliamirkhanlou@yahoo.com}, Behnam Mohammadi\footnote{be.mohammadi@urmia.ac.ir}\\
Department of Physics, Urmia University, Urmia, Iran}
\title{Analysis of $B^0_s\rightarrow \chi_{c1}(3872)\pi^+\pi^-$ decay}
\begin{document}
\maketitle
\begin{abstract}
Recently, the LHCb collaboration has analyzed the decay of
$B_s^0\rightarrow \chi_{c1}(3872)(\rightarrow J/\psi \pi^+ \pi^-)
\pi^+ \pi^-$ and reported the ratio of the branching fractions to
the $B_s^0\rightarrow \psi(2S)(\rightarrow
J/\psi\pi^+\pi^-)\pi^+\pi^-$ decay. The results of this study have
measured as a ratio of branching fractions
as{\setlength\arraycolsep{.75pt}
\begin{eqnarray}
\mathcal{R}&=&\frac{\mathcal{B}r(B_s^0\rightarrow\chi_{c1}(3872)\pi^+\pi^-)\times\mathcal{B}r(\chi_{c1}(3872)\rightarrow
J/\psi\pi^+\pi^-)}{\mathcal{B}r(B_s^0\rightarrow\psi(2S)\pi^+\pi^-)\times\mathcal{B}r(\psi(2S)\rightarrow
J/\psi\pi^+\pi^-)}\nonumber\\&=&(6.8\pm1.1\pm0.2)\times10^{-2},\nonumber
\end{eqnarray}}
and{\setlength\arraycolsep{.75pt}
\begin{eqnarray}
\mathcal{B_X}&=&\mathcal{B}r(B_s^0\rightarrow\chi_{c1}(3872)\pi^+\pi^-)\times\mathcal{B}r(\chi_{c1}(3872)\rightarrow
J/\psi\pi^+\pi^-)\nonumber\\&=&(1.6\pm0.3\pm0.1\pm0.3)\times10^{-6}.\nonumber
\end{eqnarray}}
For the first time, we calculated this branching fraction using
factorization. According to our calculations, ratio of branching
fractions to be $\mathcal{R}=(4.38\pm1.36)\times10^{-2}$ at
$\mu=m_b/2$ and the products related to branching fractions have
been estimated $\mathcal{B_X}=(1.08\pm0.62)\times10^{-6}$ at
$\mu=m_b$. The results are consistent with the experiment
reported.

\end{abstract}

\section{Introduction}
Hardonic three-body decays make up a large part of the branching
fraction for non-leptonic $B$ decays. Due to the non-trivial
kinematics, they contain a lot of information about the strong
phases of the two-body decays, and the theoretical
description of these decays is very challenging \cite{S.K1}.\\
Many interesting measurements in heavy flavour physics, both
within the framework of the standard model (SM) and beyond,
involve the decay of heavy-flavoured hadrons into final states
with neutral particles \cite{A.P1}.\\
The $B$ factories are where $B$ meson pairs are produced at the
threshold without additional particles. In such a place, full
reconstruction of one of the $B$ mesons provides sufficient
kinematic constraints for the reconstruction of undetected particles in the decay of the other $B$ meson \cite{Belle.1}.\\
QCD predictions of heavy particle decay rest on a firm theoretical
foundation \cite{E.Ba}. In recent years, the theoretical
investigation of weak decay including heavy hadrons has made
considerable progress. In the infinite limit of heavy quarks, the
decay rate coincides with that of the corresponding free quark
decay. These types of corrections have a non-perturbative origin
and are suppressed by at least two powers in the heavy quark mass \cite{I.I}.\\
Recently, exotic hadrons beyond the usual quark-model have been
observed in experiments. These hadrons are interpreted as hadron
molecular states, tetraquark states, pentaquark states, glueballs,
quark-gluon hybrids, and many others. Among the exotic hadrons is
the $\chi_{c1}$(3872) meson, which is considered as a four-quark
(tetraquark) state \cite{G.G.1}. The strong and radiative decays
have been
calculated for this meson \cite{J.L.1}.\\
The internal structure of these exotic hadrons cannot be
determined only by mass spectrum. We need to further study their
production processes or decay behaviors. In the production of
$\chi_{c1}$(3872), the $D^*D$ rescattering mechanism has a large
contribution \cite{P.A.1}. This mechanism affects the pattern of
the decay rate
of $B$ and $B_s$ hadrons. At such a time, coupling constants become especially important \cite{T.M.1}.\\
The strong coupling constants between heavy and light mesons are
among the essential components to describe low-energy hadron
interactions. Using this coupling can provide key information to
study the nature of heavy mesons. In particular, they serve as a
useful resource in investigating the final state interactions of
the $B$ meson decay. Coupling constants are needed to understand the $J/\psi$ meson production and absorption cross sections in heavy ion collisions and vertices containing charmed mesons \cite{T.M.1}.\\
Among the theories used for strong coupling is the theory with
Yukawa couplings, which describes the interaction of mesons and
constituent quarks \cite{G.G.1}. This theory is equivalent to the
theory with four-fermion interaction. In this case, the wave
function renormalization constant of the meson field should be
equal to zero. Also, the coupling $G$, which determines the
strength of the interaction of the four-fermion, should be related to the meson mass function \cite{B.R.1}.\\
In a recent LHCb paper, the first observation of $B^0_s\rightarrow
\chi_{c1}(3872)\pi^+\pi^-$ decay was reported. They have measured
the ratio of branching fraction $R$ using branching sections
$B^0_s\rightarrow (\chi_{c1}(3872)\rightarrow
J/\psi\pi^+\pi^-)\pi^+\pi^-$ and
$B^0_s\rightarrow(\psi(2S)\rightarrow
J/\psi\pi^+\pi^-)\pi^+\pi^-$.\\
The ratios of branching fraction is reported to be
$\mathcal{R}=(6.8\pm1.1\pm0.2)\times10^{-2}$. Also, the
corresponding products of branching fractions are
$\mathcal{B}r(B_s^0\rightarrow\chi_{c1}(3872)\pi^+\pi^-)\times\mathcal{B}r(\chi_{c1}(3872)\rightarrow
J/\psi\pi^+\pi^-)=(1.6\pm0.3\pm0.1\pm0.3)\times10^{-6}$ \cite{LHcb1}.\\
In this study, we have calculated the branching fractions for the
$B_s^0\rightarrow\chi_{c1}(3872)\pi^+\pi^-$,
$B_s^0\rightarrow\psi(2S)\pi^+\pi^-$, $\psi(2S)\rightarrow
J/\psi\pi^+\pi^-$ and $\chi_{c1}(3872)\rightarrow
J/\psi\pi^+\pi^-$ decays under the factorization approach and
obtained $\mathcal{R}=(4.38\pm1.36)\times10^{-2}$ at $\mu=m_b/2$.
We have estimated the products relating to branching fractions
$(1.08\pm0.62)\times10^{-6}$ at $\mu=m_b$, which is in agreement
with experimental results.

\section{Branching fractions for $B_s^0\rightarrow\chi_{c1}(3872)\pi^+\pi^-$ and
$B_s^0\rightarrow\psi(2S)\pi^+\pi^-$}

\begin{figure}[t]
\begin{center} \includegraphics[scale=1]{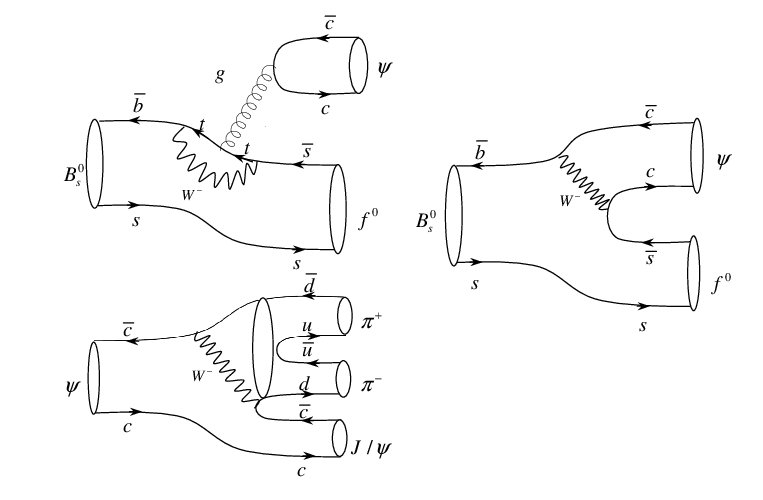}
\caption{\label{fig1} Feynman diagrams contributing to the
$B^0_s\rightarrow\psi f_0$ and $\psi\rightarrow J/\psi\pi^+\pi^-$
decays [$\psi=\psi(2S), \chi_{c1}(3872)$].}
\end{center}
\end{figure}

In the last decade, the investigation of the heavy meson sector
has gained great importance, especially for detailed tests of the
SM and exploration of physics beyond. The interactions of the
lightest hadrons, pions, with themselves as well as with kaons are
known with great precision. The combination of dispersion
relations in the form of Roy or Roy-Steiner equations, constrained
by chiral perturbation theory at the lowest energies and using
experimental data as input, has extended our knowledge of the
leading partial waves of pion-pion \cite{M.A.1}. Now here we have
the decays of $B_s^0\rightarrow\chi_{c1}(3872)\pi^+\pi^-$ and
$B_s^0\rightarrow\psi(2S)\pi^+\pi^-$. In the
$B_s^0\rightarrow\psi\pi^+\pi^-$ decay, where $B^0_s$ decays into
$\psi=\psi(2S), \chi_{c1}(3872)$ and a
pair of $\pi^+\pi^-$ light pseudoscalar hadrons.\\
In these decays, the pion-pion scattering process is dominant in
the $f^0$(980) resonance region \cite{M.A.2}. In such a case, an
S-wave pion pair is produced from a quark-antiquark pair, and the
interactions of the final state are described by the scalar form
factor (the idea of the scalar-source model) \cite{W.H.1}. In
contrast to dynamical resonance generation models, the scalar
resonances are considered as $q\bar{q}$ or tetraquark states. If
$B^0_s$ is above the $f_0\psi$ threshold, the width of
$B_s^0\rightarrow\psi\pi^+\pi^-$ can be written \cite{C.M.1}
\begin{eqnarray}\label{eq1}
\Gamma(B_s^0\rightarrow\psi\pi^+\pi^-)=2\Gamma(B_s^0\rightarrow\psi
f_0)\mathcal{B}r(f_0\rightarrow\pi^+\pi^-).
\end{eqnarray}
We have
$\Gamma(f_0\rightarrow\pi^+\pi^-)=34.2^{+13.90}_{-11.80}(stat)^{+8.80}_{-2.50}(syst)$
MeV \cite{Belle11}, and the width $\Gamma(B_s^0\rightarrow\psi
f_0)$ can be easily obtained from the naive factorization
approach. According to Fig. \ref{fig1},
the decays amplitude is shown
\begin{eqnarray}\label{eq2}
\mathcal{M}(B_s^0\rightarrow\psi
f_0)=iG_{F}m_{\psi}f_{\psi}\epsilon_{\psi}\cdot p_{B_s}
F_1^{B_s\rightarrow
f_0}(m_{\psi}^2)\Big(V_{cb}V^*_{cs}a_2-V_{tb}V^*_{ts}(a_3+a_9+r_{\chi}^{\psi}(a_5+a_7))\Big),
\end{eqnarray}
where $\epsilon_{\psi}\cdot p_{B_s}=|p_c|m_{\psi}/m_{B_s}$,
$p_c=\sqrt{(m^2_{B_s}-(m_{\psi}+m_{f_0})^2)(m^2_{B_s}-(m_{\psi}-m_{f_0})^2)}/2m_{B_s}$
\cite{CeM} and
$r_\chi^{\psi}=(2m_{\psi}/m_b)(f_{\psi}^{\bot}/f_{\psi})$
\cite{D.Me}. $f_0$ meson is
$(u\bar{u}+d\bar{d})/\sqrt{2}+s\bar{s}$ \cite{PDG1}. To achieve
the form factor $F_1$ we take the form \cite{D.Me}
\begin{eqnarray}\label{eq3}
F_1(q^2)=\frac{m_{f_0}+m_{B_s}}{2\sqrt{m_{f_0}m_{B_s}}}\xi(\omega),
\end{eqnarray}
with $\xi(\omega)=1-\rho^2(\omega-1)$, $\rho^2=1.2^{+7}_{-3}$
\cite{R.D1} and
$\omega=(m_{B_s}^2-m_{\psi}^2+m_{f_0}^2)/(2m_{B_s}m_{\psi})$. The
$\xi(\omega)$ and $\rho$ are Isgur-Wise function and the slope
parameter respectively \cite{MN1}. $G_F=(1.16639\pm0.00001)\times
10^{-5}$, $V_{pb}V_{ps}^* (p= c,t)$ the Cabibbo-Kobayashi-Maskawa
(CKM) matrix elements \cite{HFLAV2} and $a_j=c_j+ c_{j\pm1}/3$
(j=1, \ldots, 10), for j=odd(even) is composed of the effective
Wilson coefficients
$c_j$ at the renormalization scales of $\mu= 2m_b, m_b, m_b/2$ defined in \cite{M.Be1}.\\
The branching fractions of the $B^0_s\rightarrow\psi(2S)f_0$ and
$B^0_s\rightarrow\chi_{c1}(3872)f_0$ decays under the
factorization approach are given by
\begin{eqnarray}\label{eq4}
\mathcal{B}r(B^0_s\rightarrow\psi f_0)=\frac{\Gamma
(B^0_s\rightarrow\psi f_0)}{\Gamma_{tot}},
\end{eqnarray}
where the $\Gamma_{tot}$ for $B_s^0$ meson is
$(4.33\pm0.01)\times10^{-13}$ GeV \cite{PDG1}.

\section{Branching fractions for $\psi(2S)\rightarrow J/\psi\pi^+\pi^-$ and $\chi_{c1}\rightarrow J/\psi\pi^+\pi^-$}

In phenomenology, the vertices of charmed mesons play an important
role in meson interactions. They help us investigate the final
state interactions in hadronic $B$ decays.
The charmed mesons are considered intermediate states in the branching ratios for non-leptonic $B$ meson decays \cite{S.M.1}.\\
On the other hand, strong couplings between charm mesons and other
hadrons can help us to study the production of $J/\psi$. The ATLAS
collaboration observed the $X$(3872) meson while measuring the
cross-section of prompt and non-prompt $\psi(2S)$ meson production
in the $J/\psi\pi^+\pi^-$ decay channel with 2011 data at the
center-of-mass energy ($\sqrt{s}=7TeV$) \cite{ATLAS1}.\\
The $J/\psi\pi^+\pi^-$ final state allows good invariant mass
resolution through the use of the constrained fit and provides a
straightforward way to compare the production characteristics of
the $\psi(2S)$ and $X$(3872) states that are nearly close in
mass \cite{COM1}.\\
Based on a recent combined data analysis, the decays
$X(3872)\rightarrow J/\psi\rho\rightarrow J/\psi\pi^+\pi^-$ and
$X(3872)\rightarrow J/\psi\omega\rightarrow J/\psi\pi^+\pi^-\pi^0$ have almost the same branching fractions \cite{ATLAS1}.\\
The Fig. \ref{fig1} analysis shows the contribution from the
$[\pi^+\pi^-]$ channel resonances (intermediate $d\bar{d}$). The
matrix element $\langle\pi^+\pi^-|\bar{d}d|0\rangle$ receives a
contribution from vector and scalar resonances, which include
$V:\rho(770),
\rho(1690), \omega(782)$ and $S:f_0(980), f_0(2050)$.\\
According to particle data group measurements \cite{PDG1}, it is
$\psi\rightarrow\omega J/\psi\rightarrow J/\psi\pi^+\pi^-$. To
calculate $\psi\rightarrow J/\psi\pi^+\pi^-$ using Eq.
(\ref{eq1}), we need to calculate $\psi\rightarrow\omega J/\psi$.
We have \cite{M.A.1}
\begin{eqnarray}\label{eq5}
\Gamma(\psi\rightarrow
J/\psi\omega)=\frac{g^2}{96\pi}\frac{\lambda^{3/2}(m^2_{\psi},m^2_{\omega},m^2_{J/\psi})}{m^3_{\psi}},
\end{eqnarray}
where $\lambda(x, y, z)=x^2+y^2+z^2-2(xy+yz+zx)$ and
$g=f_{\omega}g_{\psi\omega J/\psi}/\sqrt{2m_{\psi}m_{\omega}}$
\cite{B.E.1}. $f_{\omega}$ and $g_{{\psi\omega J/\psi}}$ are decay
constant and strong coupling constant, respectively. We have the
value of $\mathcal{B}r(\omega\rightarrow\pi^+\pi^-)=(1.53^{+0.11}
_{-0.13})\%$ \cite{PDG1}. We obtained $g_{{\psi\omega
J/\psi}}=44.65$ MeV  and calculated the branching fraction
$\chi_{c1}(3872)\rightarrow J\psi\pi^+\pi^-$ and
$\psi(2S)\rightarrow J\psi\pi^+\pi^-$ decays. The $\Gamma_{tot}$
for $\chi_{c1}(3872)$ and $\psi(2S)$ meson is $1.19\pm0.21$ and
$(294\pm8)\times10^{-6}$ MeV respectively \cite{PDG1}.

\section{Branching fractions $\mathcal{R}$ and $\mathcal{B_X}$}

The LHCb collaboration, the first observation of the
$B^0_s\rightarrow(\chi_{c1}(3872)\rightarrow
J/\psi\pi^+\pi^-)\pi^+\pi^-$ decay, have reported. Their results
were measured in the form of a ratio of branching fractions using
the $B^0_s\rightarrow\psi(2S)\pi^+\pi^-$ decay as a normalization
channel. The $\mathcal{R}$
ratio and the product of branching fractions $\mathcal{B_X}$ have been obtained with the following formulas \cite{LHcb1}: \\
\begin{eqnarray}\label{eq6}
&&\mathcal{R}=\frac{\mathcal{B}r(B_s^0\rightarrow\chi_{c1}(3872)\pi^+\pi^-)\times\mathcal{B}r(\chi_{c1}(3872)\rightarrow
J/\psi\pi^+\pi^-)}{\mathcal{B}r(B_s^0\rightarrow\psi(2S)\pi^+\pi^-)\times\mathcal{B}r(\psi(2S)\rightarrow
J/\psi\pi^+\pi^-)},\nonumber\\
&&\mathcal{B_{X}}=\mathcal{B}r(B_s^0\rightarrow\chi_{c1}(3872)\pi^+\pi^-)\times\mathcal{B}r(\chi_{c1}(3872)\rightarrow
J/\psi\pi^+\pi^-).
\end{eqnarray}
We calculated the value of $\mathcal{R}$ and $\mathcal{B_X}$ using
the branching ratio that we estimated. We obtained the value of
$\mathcal{R}$ in three different scales and calculated the value
of $\mathcal{B_X}$ using it. In process related to
$B_s^0\rightarrow\psi\pi^+\pi^-$, $\psi$ can be formed as an
intermediate resonance particle and subsequently it can decay into
$J/\psi\pi^+\pi^-$. The $B^0_s\rightarrow
J/\psi\pi^+\pi^-\pi^+\pi^-$ decay, where the $J/\psi\pi^+\pi^-$
combination does not originate from a $\chi_{c1}$(3872) meson.
Therefore parameterised by the product of the $B^0_s$ and the
phase-space function describing three-body combinations from
five-body decays.

\section{Numerical results and conclusion}

The numerical values of meson masses and decay constants are
tabulated in Tab. \ref{tab1}.
\begin{table}[t]
\centering\caption{\label{tab1} The meson masses and decay
constants (in MeV) \cite{PDG1}}
\begin{tabular}{|c c c c c|}
  \hline
  $m_{B^0_s}$  &  $m_{\chi_{c1}(3872)}$& $\psi(2S)$ & $m_{J/\psi}$ & $m_{\pi^{\pm}}$
\\\hline
$5366.92\pm0.10$ & $3871.65\pm0.06$ & $3686.10\pm0.06$ &
$3096.900\pm0.006$ & $139.57039\pm0.00018$
\\\hline
 $m_{f_0}$ & $m_{\omega}$ & $m_{b}$ &  & \\\hline
$990\pm20$ & $782.66\pm0.13$ & $4180^{+30}_{-20}$ &  &
\\\hline\hline
 $f_{\psi}$ & $f_{\psi}^{\bot}$ \cite{F.M},\cite{P.B} & $f_{\omega}$ \cite{P.B.1} & $f_{\chi_{c1}}$ & \\\hline
 $282\pm14$ & $255\pm33$ & $195\pm3$ & $234\pm52$  & \\\hline
\end{tabular}
\end{table}
The elements of the CKM matrix used in the calculations have the
following values:\\
\begin{tabular}{cccc}
$V_{cs}$ & $V_{cb}$ & $V_{tb}$ & $V_{ts}$\\
$0.975\pm0.006$ & $(40.8\pm1.4)\times10^{-3}$ & $1.014\pm0.029$ & $(41.5\pm0.9)\times10^{-3}$ \\
\end{tabular}\\
Decays of beauty hadrons to final states with charmonia are ideal
for studying as an investigation. We determined the decay rates of
$B^0_s\rightarrow \psi(2S)\pi^+\pi^-$, $\psi(2S)\rightarrow
J/\psi\pi^+\pi^-$,$B^0_s\rightarrow\chi_{c1}\pi^+\pi^-$ and
$\chi_{c1}\rightarrow J/\psi\pi^+\pi^-$, using Feynman diagrams,
the naive factorization approach and channel resonance. Then we
estimated branching fractions using them. The results by comparing the experimental value are presented in Tab. \ref{tab3}.\\
\begin{table}[t]
\centering\caption{\label{tab3} The numerical result of branching
ratio for the $B^0_s\rightarrow \psi\pi^+\pi^-$, $\psi\rightarrow
J/\psi\pi^+\pi^-$, the ratio of branching fractions $\mathcal{R}$
and product of branching fractions $\mathcal{B_X}$.}
\begin{tabular}{|ccccc|}
  \hline
 decay mode &  & $\mu=m_c$ &  & Exp. $(\times10^{-2})$ \cite{PDG1} \\
\hline $\mathcal{B}r(\chi_{c1}(3872)\rightarrow J/\psi\pi^+\pi^-)$
& & $3.10\pm0.66$ &  & $3.80\pm1.20$\\\hline
$\mathcal{B}r(\psi(2S)\rightarrow J/\psi\pi^+\pi^-)$ &
 & $34.01\pm0.90$ &  & $34.68\pm0.30$
\\\hline\hline
 & $\mu=m_b/2$ & $\mu=m_b$ & $\mu=2m_b$ & Exp. $(\times10^{-5})$ \cite{PDG1} \\
\hline $\mathcal{B}r(B^0_s\rightarrow \chi_{c1}(3872)\pi^+\pi^-)$
& $1.28\pm0.72$ & $3.47\pm2.93$ & $6.01\pm2.30$  & -
\\\hline $\mathcal{B}r(B^0_s\rightarrow\psi(2S)\pi^+\pi^-)$ &
$2.66\pm1.38$ & $7.75\pm3.73$ & $13.67\pm6.46$ & $6.90\pm1.20$
\\\hline\hline
 & $\mu=m_b/2$ & $\mu=m_b$ & $\mu=2m_b$ & Exp. $(\times10^{-2})$ \cite{LHcb1} \\
\hline $\mathcal{R}$ & $4.38\pm1.36$ & $4.08\pm1.37$ &
$4.01\pm2.12$ & $6.80\pm1.10\pm0.02$\\\hline\hline
 & $\mu=m_b/2$ & $\mu=m_b$ & $\mu=2m_b$ & Exp. $(\times10^{-6})$ \cite{LHcb1} \\
\hline $\mathcal{B_X}$ & $0.40\pm0.18$ & $1.08\pm0.62$ &
$1.86\pm1.14$ & $1.60\pm0.30\pm0.10\pm0.30$
\\\hline
\end{tabular}
\end{table}
We performed calculations $\psi(2S)\rightarrow J/\psi\pi^+\pi^-$
and $\chi_{c1}(3872)\rightarrow J/\psi\pi^+\pi^-$ at $\mu=m_c$
scale. Also, we calculated the
$B^0_s\rightarrow\chi_{c1}(3872)\pi^+\pi^-$ and $B^0_s\rightarrow
\psi(2S)\pi^+\pi^-$ decays on three scales of $m_b$. In this way,
we obtained the values of $\mathcal{R}$ and $\mathcal{B_X}$ in these scales.\\
The $\psi(2S)$ and $\chi_{c1}$(3872) mesons are nearly close in
mass. As a result, the decay of
$B^0_s\rightarrow\chi_{c1}(3872)\pi^+\pi^-$ and $B^0_s\rightarrow
\psi(2S)\pi^+\pi^-$ have almost the same
branching fractions.\\
In the past decade, many new excited heavy meson states have been
discovered experimentally. Many theoretical studies have been
conducted to identify their properties and structures. However, it
is still difficult to determine the structure of these mesons.\\
Also, this decay is one of the strong decays and is of the type of
decay of charm hadron meson into two light heavy mesons, and
despite the fact that this decay has current-current diagram, this
method has not been successful in solving this decay.\\
With recent investigations of $B$ factories and the planned
emphasis on heavy flavor physics in future experiments, the role
of $B$ decay in providing fundamental tests of the standard model
and potential effects of new physics will continue to grow.\\
Our work provides a precise framework for evaluating strong
interactions for a large classification of three-body non-leptonic
$B$ decays. In the case of branching ratios, theoretical
predictions have large uncertainties due to the hadronic
distributions, the hard scattering, and the renormalization scales
in factorizable amplitudes. In the heavy quark limit, matrix
elements can be expressed in terms of certain non-perturbative
input values such as transition form factors.\\
The power corrections beyond the heavy quark limit generally
introduce large theoretical uncertainties. Also, the CKM factors
mainly provide a general factor for the branching ratios and do
not introduce many uncertainties to the numerical results.


\begin{thebibliography}{20}
\bibitem{S.K1}
S. Krankl, T. Mannel and J. Virto, \textit{Three-body non-leptonic
$B$ decays and QCD factorization}, Nucl. Phys. B \textbf{899}
(2015) 247.
\bibitem{A.P1}
A. Poluektov and A. Morris, \textit{Oscillations of $B^0_s$ mesons
as a probe of decays with unreconstructed particles}, JHEP
\textbf{02} (2020) 163.
\bibitem{Belle.1}
C.L. Hsu et al., Belle collaboration, \textit{Search for $B^0$
decays to invisible final states at Belle}, Phys. Rev. D
\textbf{86} (2012) 032002.
\bibitem{E.Ba}
E. Bagan, P. Ball, V.M. Braun and P. Gosdzinsky,
\textit{Theoretical update of the semileptonic branching ratio of
$B$ mesons}, Phys. Lett. B \textbf{342} (1995) 362.
\bibitem{I.I}
I.I. Bigi, N. Uraltsev and A. Vainshtein, \textit{Non-perturbative
corrections to inclusive beauty and charm decays : QCD versus
phenomenological models}, Phys. Lett. B \textbf{293} (1995) 430.
\bibitem{G.G.1}
G. Ganbold, T. Gutsche, M.A. Ivanov and V.E. Lyubovitskij,
\textit{On the meson mass spectrum in the covariant confined quark
model}, J. Phys. G: Nucl. Part. Phys \textbf{42} (2015) 075002.
\bibitem{J.L.1}
J. Lu, G.L. Yu and ZH.G. Wang, \textit{The strong vertices of
charmed mesons $D, D^*$ and charmonia $J/\psi, \eta_c$}, Eur.
Phys. J. A \textbf{59} (2023) 195.
\bibitem{P.A.1}
P. Artoisenet, E. Braaten and D. Kang, \textit{Using line shapes
to discriminate between binding mechanisms for the $X$(3872)},
Phys. Rev. D \textbf{82} (2010) 014013.
\bibitem{T.M.1}
T.M. Aliev and K. Simsek, \textit{Strong coupling constants of
charmed and bottom mesons with light vector mesons in QCD sum
rules}, Phys. Rev. D \textbf{104} (2021) 074034.
\bibitem{B.R.1}
B. Rosenstein, B.J. Warr and S.H. Park, \textit{The four fermi
theory is renormalizable in (2+1)-dimensions}, Phys. Rev. Lett
\textbf{62} (1989) 1433.
\bibitem{LHcb1}
R. Aaij et al., LHCb collaboration, \textit{Observation of the
$B^0_s\rightarrow\chi_{c1}(3872)\pi^+\pi^-$ decay}, JHEP
\textbf{07} (2023) 084.
\bibitem{M.A.1}
M. Albaladejo, J.T. Daub, C. Hanhart, B. Kubis and B. Moussallam,
\textit{How to employ $\bar{B}^0_d\rightarrow J/\psi(\pi\eta,
K\bar{K}$) decays to extract information on $\pi\eta$ scattering},
JHEP \textbf{04} (2017) 010.
\bibitem{M.A.2}
J.T. Daub, C. Hanhart and B. Kubis, \textit{A model-independent
analysis of final-state interactions in
$\bar{B}^0_{d/s}\rightarrow J/\psi\pi\pi$}, JHEP \textbf{02}
(2016) 009.
\bibitem{W.H.1}
W.H. Liang and E. Oset, \textit{$B^0$ and $B^0_s$ decays into
$J/\psi f_0$(980) and $J/\psi f_0$(500) and the nature of the
scalar resonances}, Phys. Lett. B \textbf{737} (2014) 70.
\bibitem{C.M.1}
C. Meng and K.T. Chao, \textit{Decays of the X(3872) and
$\chi_{c1}(2P)$ charmonium}, Phys. Rev. D \textbf{75} (2007)
114002.
\bibitem{Belle11}
T. Mori et al., Belle collaboration, \textit{High statistics study
of the $f_0$ (980) resonance in
$\gamma\gamma\rightarrow\pi^+\pi^-$ production}, Phys. Rev. D
\textbf{75} (2007) 051101.
\bibitem{CeM}
C. Meng and K.T. Chao, \textit{Decays of the $X(3872)$ and
$\chi_{c1}(2P)$ charmonium}, Phys. Rev. D \textbf{75} (2007)
114002.
\bibitem{D.Me}
D. Melikhov and B. Stech, \textit{Weak form factors for heavy
meson decays: An update}, Phys. Rev. D \textbf{62} (2000) 014006.
\bibitem{PDG1}
R.L. Workman et al., Particle Data Group, \textit{Review of
particle physics (2022}, Prog. Theor. Exp. Phys \textbf{2022}
(2022) 083C01.
\bibitem{R.D1}
R.D Kenway, \textit{The Isgur-Wise function}, Nucl. Phys. Proc.
Suppl \textbf{34} (1994) 153.
\bibitem{MN1}
M. Neubert and V. Rieckert, \textit{The Isgur-Wise function from
the lattice}, Nucl. Phys. B \textbf{97} (1992) 382.
\bibitem{HFLAV2}
Y. Amhis et al., Heavy Flavor Averaging Group, \textit{Averages of
$b$-hadron, $c$-hadron, and $\tau$-lepton properties as of summer
2016}, Eur. Phys. J. C \textbf{77} (2017) 895.
\bibitem{M.Be1}
M. Beneke, G. Buchalla, M. Neubert and C.T. Sachrajda, \textit{QCD
factorization in $B\rightarrow \pi K, \pi\pi$ decays and
extraction of wolfenstein parameters}, Nucl. Phys. B \textbf{606}
(2001) 245.
\bibitem{S.M.1}
S. Momeni and M. Saghebfar, \textit{Charm meson couplings in
hard-wall Holographic QCD}, Eur. Phys. J. C \textbf{81} (2021)
102.
\bibitem{ATLAS1}
M. Aaboud et al., ATLAS collaboration, \textit{Measurements of
$\psi(2S)$ and $X(3872)\rightarrow J/\psi\pi^+\pi^-$ production in
pp collisions at $\sqrt{s}=8$ TeV with the ATLAS detector}, JHEP
\textbf{01} (2017) 117.
\bibitem{COM1}
M. Aghasyan et al., COMPASS Collaboration, \textit{Search for
muoproduction of $X$(3872) at COMPASS and indication of a new
state \~{X}(3872)}, Phys. Lett. B \textbf{783} (2018) 334.
\bibitem{B.E.1}
B. El-Bennicha and F.E. Serna, \textit{Heavy-meson chiral
Lagrangians, effective flavored couplings, SU(4) flavor breaking
and their consequences}, PoS. CHARM \textbf{2020} (2021) 025.
\bibitem{F.M}
F.M. Al-Shamali and A.N. Kamal, \textit{Nonfactorization and final
state interactions in $(B,B_s)\rightarrow\psi P$ and $\psi V$
decays}, Eur. Phys. J. C \textbf{4} (1998) 669.
\bibitem{P.B}
P. Ball and V.M. Braun, \textit{The $\rho$ meson light-cone
distribution amplitudes of leading twist revisited}, Phys. Rev. D
\textbf{54} (1996) 2182.
\bibitem{P.B.1}
P. Ball and R. Zwicky, \textit{$B_{d,s}\rightarrow\rho, \omega,
K^*,\phi$ decay form factors from light-cone sum rules revisited},
Phys. Rev. D \textbf{71} (2005) 014029.
\end{thebibliography}
\end{document}